\documentclass[doublespace,onecolumn]{SSA-SRL}

\usepackage{color}
\definecolor{txtcol}{rgb}{0.,0.,0.}

\citethisauthor{Shcherbakov, R.}
\doi{10.1785/0220240157}
\recdate{03 October 2024}

\begin{document}

\title{A Stochastic Model for Induced Seismicity at the Geothermal Systems: A Case of the Geysers}

\author[*1,2\orc{0000-0002-3057-0851}]{Robert Shcherbakov}

\affil[1]{Department of Earth Sciences, Western University, London, Ontario, N6A 5B7, Canada}{\auorc[https:// orcid.org/]{0000-0002-3057-0851}{(R.S.)}}
\affil[2]{Department of Physics and Astronomy, Western University, London, Ontario, N6A 3K7, Canada}{}
\corau{*Corresponding author: rshcherb@uwo.ca}

\begin{abstract}
Induced seismicity has emerged as a source of a significant earthquake hazard associated with recent development of unconventional energy resources. Therefore, it is imperative to develop stochastic models that can accurately describe the observed seismicity rate and forecast its evolution. In this study, a mechanism suggested by linear response theory is incorporated into a stochastic earthquake model to account for changes in the seismicity rate. It is derived that the induced rate can be modelled as a convolution of the forcing, related to fluid injection operations, and a specific response kernel. The model is incorporated into a Bayesian framework to compute the probabilities for the occurrence of the largest expected events during future time intervals. The applicability of the model is illustrated by analyzing the injection and seismicity data at the Geysers geothermal field in California. The suggested approach provides further insight into the probabilistic assessment of earthquake hazard associated with fluid injection operations. It also can be used for probing the rheological properties of the subsurface by analysing the inherent characteristic time-scales associated with the subsurface response to external forcing.
\end{abstract}

\maketitle

\section{Introduction}

The occurrence of earthquakes is driven by complex physical processes operating across vast temporal and spatial scales within the Earth's brittle crust \citep{KanamoriB04a,Ben-Zion08a}. Tectonic seismicity primarily arises from the gradual loading, deformation and/or collision of the continental and oceanic plates. It is characterized by a relatively moderate rate of the background activity, occasionally punctuated by the occurrence of large earthquakes, which often trigger aftershock sequences. However, earthquakes can also be induced by various anthropogenic activities, where energy and/or resource extraction operations can lead to the redistribution of subsurface stresses and alterations in pore pressure, thereby precipitating seismic events \citep{Ellsworth13a,GrigoliCPR17a,SchultzSBEBE20a}. Therefore, it is crucial to recognize and integrate such anthropogenic factors into the quantitative models that underpin our understanding of earthquake processes. This is also critical for any seismic hazard mitigation strategies to minimize the risk associated with induced seismicity \citep{McGarr2014a,GoebelB18a,LeeEGTG2019a,HagerDFJN2021a,RitzRW2022a}. 

Several physics based mechanisms have been proposed to explain the occurrence of induced seismicity. For subsurface operations involving fluid injection, it was suggested that the seismicity is driven by the advancing fluid diffusion front \citep{ShapiroD09a,BachmannWGW12a}. The corresponding increase in pore pressure results in the decrease of the effective normal stress that can lead to slippage on preexisting planes of weakness. However, pore pressure diffusion may not play a dominant role when explaining seismicity in some instances of hydraulic fracturing operations or after the shut-in of an injection well. In such cases, it was shown that the transfer of poroelastic stresses can trigger induced seismicity \citep{BaoE16a,AtkinsonEGW16a}. It was also suggested that induced seismicity can be triggered farther away from injection sites by propagating aseismic slip that may outpace pore pressure diffusion front \citep{GuglielmiCAHE2015a,BhattacharyaV19a,Wynants-MorelCBA2020a,YeoBGL2020a}. 

In thermodynamic considerations of the out-of-equilibrium systems, macroscopic theory of irreversible processes plays a critical role. Within it, linear response theory is used to quantify the effects of fluctuations and/or external forcing on the time evolution of nonequilibrium systems \citep{LiviP2017a}. In many cases, the response of such systems to forcing is quantified by the presence of the spectrum of characteristic time scales which are fundamental to the behaviour and physics of these nonequilibrium systems \citep{HasselmannHGOS1997a,Lucarini2018a}. Seismogenic regions are examples of such nonequilibrium systems. Therefore, the rate of fluid injection plays a role of external forcing and the occurrence of induced earthquakes represents the response of the seimsmogenic region to such forcing. Therefore, I hypothesize that the observed induced seismicity rate can be quantified using the methods of liner response theory.

Statistical seismology treats the occurrence of earthquakes as a stochastic point process in space, time and magnitude domains \citep{Vere-Jones10a}. The most successful point process to date to approximate the occurrence of earthquakes is the Epidemic Type Aftershock Sequence (ETAS) model \citep{Ogata88a,Ogata99a}. It captures two most important ingredients of tectonic seismicity, i.e. the occurrence of background events and triggering of aftershock sequences. In several studies, the standard ETAS model was used to approximate the rate of induced seismicity in Oklahoma and Arkansas \citep{LlenosM13a,AochiMG21a}, at the Salton Sea geothermal site \citep{BrodskyL13a,LlenosM16a}, in Alberta, Canada \citep{KothariSA20a} and mining induced seismicity \citep{SedghizadehSB2024a}. However, this model needs further generalization when it comes to describing main driving factors that lead to the occurrence of induced earthquakes.

\begin{figure}[!ht]
\centering
{\includegraphics*[width=.9\textwidth, scale=0.75, viewport= 15mm 35mm 210mm 250mm]{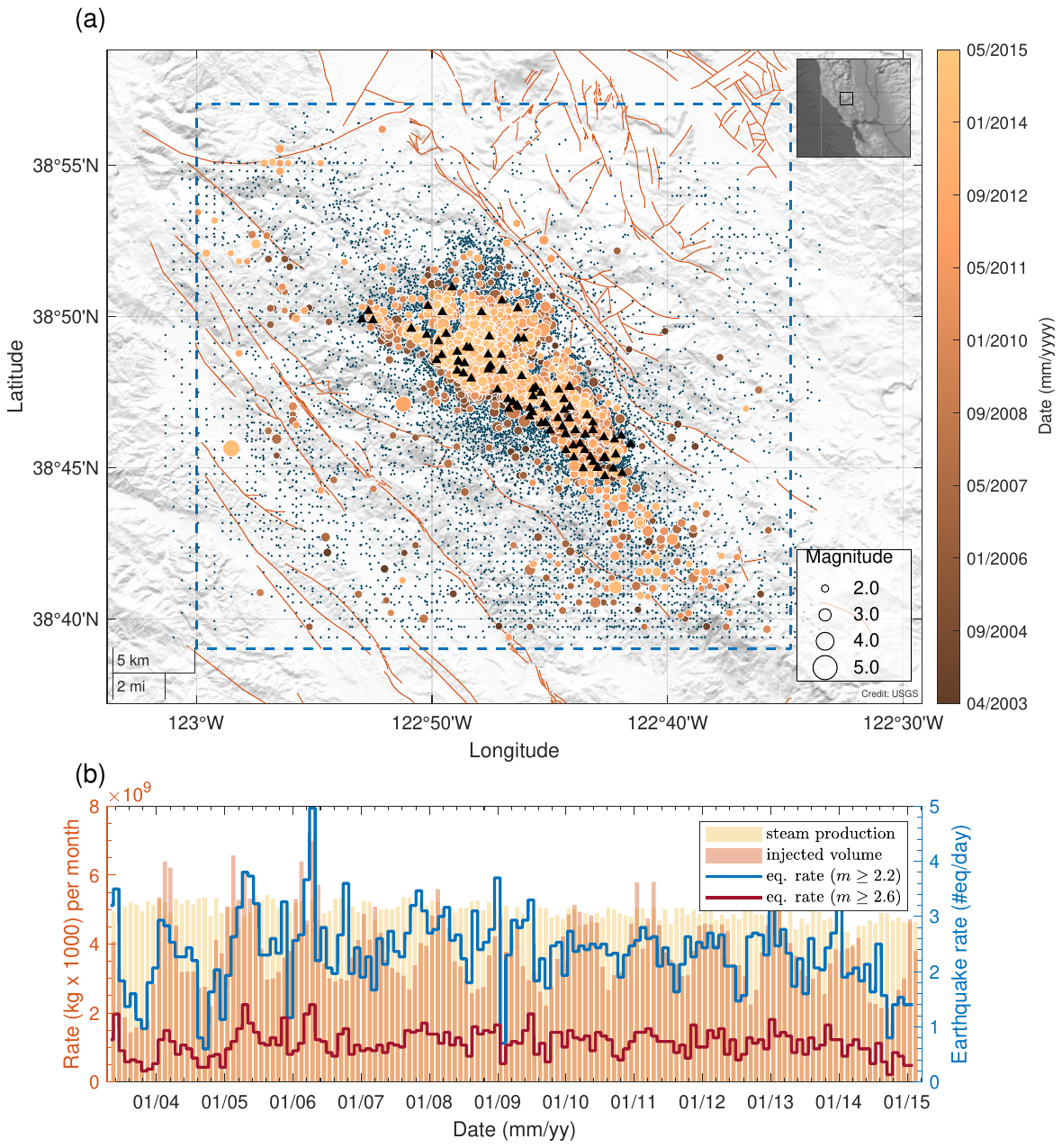}}
\caption{Spatial distribution of seismicity, earthquake rates, fluid injection and steam production monthly volumes at the Geysers geothermal field. (a) The events above magnitude $2.6$ are plotted as colored solid circles with varying radii reflecting their magnitudes. The blue dots indicate all other events below magnitude $2.6$. Brown line segments plot the quaternary faults. The black solid triangles show the active injection wells. (b) The monthly earthquake rates are computed for events above magnitudes $m\ge 2.2$ (blue line) and $m\ge 2.6$ (dark red line). Monthly fluid injection and steam production volumes are given as colored bars.}
\label{fig1}
\end{figure}

To model the induced seismicity rate, the most straightforward approach is to assume that it is directly proportional to the fluid injection rate. This was used in modelling the induced seismicity associated with earthquake swarms at NW Bohemia \citep{HainzlO05a}, the Basel geothermal experiment \citep{BachmannWWH11a,BroccardoMWSG17a}, the fluid injection in Oklahoma \citep{LangenbruchWZ2018a}, the Val d’Agri oil field in Italy \citep{ImprotaVPC2015a}, the Geysers geothermal site in California \citep{HoltzmanPPWR2018a,Garcia-Aristizabal2018a,SaezL2023a}, the Preston New Road (United Kingdom) hydraulic fracturing operations \citep{ClarkeVKBK2019a,ManciniWSB2021a}. These observations indicate that the control of fluid injection rates may be used to manage the intensity of induced seismicity and the occurrence of the largest events \citep{DempseyR2019a}. The use of the convolution operation and Omori like decay kernel was used in the study of the Otaniemi geothermal reservoir stimulation \citep{KimA2023a}. From an earthquake hazard assessment point of view, the magnitude of the largest event is of special concern and several approaches were suggested to constrain a possible maximum induced magnitude \citep{McGarr2014a,vanderElstPWGH2016a,LangenbruchWZ2018a,ZollerH2023a,DempseyS2023a}.

The induced seismicity rate has been related to the changes in the stressing rate due to redistribution of poroelastic or thermal stresses and fluid pressure \citep{WeingartenGGBR2015a,AlghannamJ20a}. This was demonstrated in the context of fully-coupled poroelastic geomechanical modeling and assumption of a rate-and-state formalism \citep{SegallL15a,DengLH16a}. The rate-and-state formulation was also used to model and forecast the seismicity rate in Oklahoma and Arkansas \citep{NorbeckR18a,ZhaiSMC2019a}. Geomechanical modeling plays an important role in understanding the physical mechanisms of induced seismicity but it is also limited by the stochastic nature of the occurrence of earthquakes and heterogeneous properties and structure of the subsurface. 

In this work, I consider a generalization of the temporal ETAS model to account for the induced aspects of fluid injection operations. This is done in the framework of linear response theory. I assume that the background rate is no longer constant but is given as a convolution operation between the fluid injection rate with a specific kernel function. The model is applied to one example of induced seismicity associated with the geothermal system in California, the Geysers (Fig.~\ref{fig1}a). This site represents one important example of a geothermal power generating facility that has been operational since 1960s. It is the largest such system in the world and generates approximately 850 MW of electricity per year. The site is prone to a high level of seismic activity which is modulated by subsurface injection of water to produce steam that is used to generate electricity \citep{MajerP2007a}. There are more than 70 injection wells reaching up to 5 km depth \citep{HartlineWW2019a}. Water injection volumes vary significantly throughout the year with more water injected during winter months (Fig.~\ref{fig1}b). The induced seismicity is primarily associated with water injection rather than steam production, and it is most likely driven by changes in thermal stresses related to thermal expansion and contraction of subsurface rocks \citep{Martinez-GarzonKSBDH2014a,Martinez-GarzonKBD16a,HoltzmanPPWR2018a,HartlineWW2019a,GrittoJHU2023a}.

\section{A stochastic earthquake rate model}\label{models}

The original ETAS model was introduced to approximate tectonic seismicity as a stochastic point process in time and space \citep{Ogata88a,Ogata99a}. It incorporates two mechanisms that are responsible for the occurrence of earthquakes. The background events occur randomly with a constant or stationary spatially varying rate and follow Poisson statistics. Typically, this mechanism is associated with slow tectonic loading and random occurrence of events. To model the triggering aspects of seismicity, each event is capable of producing offspring events with the rate that decays hyperbolically in time and space. As a result, the total rate in the ETAS model at given time is a superposition of the background rate and a contribution from each past event.

To model the seismicity rate observed in the regions associated with anthropogenic energy related activities, one has to account for the inducing aspects of such activities. This can be accomplished by modifying the background term $\mu$ in the ETAS model and assuming that it is no longer stationary but depends on the rate of anthropogenic activities. In what follows, I assume that the anthropogenic activity is proportional to the variations in the rate of fluid injection volume or pressure. Therefore, the original ETAS model \citep{Ogata88a} can be reformulated in a more general form
\begin{equation}\label{etas_induced}
    \lambda_\omega(t|\mathcal{H}_t,\mathcal{F}_t) = \mu_\mathrm{ind}(t|\mathcal{F}_t)
    + K\,\sum\limits_{i:t_i<t}^{N_t} \frac{\mathrm{e}^{\alpha(m_i-m_0)}}{\left(\frac{t-t_i}{c}+1\right)^p}\,,
\end{equation}
where $\mu_\mathrm{ind}(t,\mathcal{F}_t)$ specifies the contribution to the rate from the background and/or other inducing processes that can trigger subsequent earthquakes and can depend on a past history of anthropogenic activities, $\mathcal{F}_t$. $m_0$ is a reference magnitude and $\{K,\,c,\,p,\,\alpha\}$ is a set of the model parameters associated with the aftershock triggering mechanism. The sum runs over the history, $\mathcal{H}_t$, of past $N_t$ events $\{m_i\},\, i=1,\ldots,N_t$, during the interval $[T_0,\,t[$ and above magnitude $m_0$ with $T_0$ being the initial start time typically set to zero.

For the functional form of the inducing term $\mu_\mathrm{ind}(t,\mathcal{F}_t)$, following linear response theory, I assume that the effect of the subsurface injection is defined as a convolution of the normalized fluid injection rate with a given kernel (see Supplemental Material):
\begin{equation}\label{rate_conv}
    \mu_\mathrm{ind}(t,\mathcal{F}_t) = \mu_0\int\limits_{T_0}^{t} G(t-t')\,F(t')\,dt'\,,
\end{equation}
where $F(t)$ specifies the forcing rate due to the processes associated with fluid injection. $\mu_0$ plays the role of a normalization factor that relates the fluid injection rate to a seismicity rate and aids the convergence of the optimization algorithm when estimating the model parameters. The kernel (response function) $G(t)$ can be defined using several functional forms that reflect the response of the subsurface media to the fluid injection and changes in the stress field. Assuming that the transient response is governed by Maxwell rheology, the response function can be specified as an exponential function \citep{HasselmannHGOS1997a}:
\begin{equation}\label{expkernel}
    G(t) = \frac{1}{t_a}e^{-\frac{t}{t_a}}\,,
\end{equation}
where $t_a$ is a single characteristic time scale intrinsic to the physical response of the subsurface. Another possible form for the kernel is to use a hyperbolic function:
\begin{equation}\label{powerlawkernel}
    G(t) = \frac{1}{\left(1 + \frac{t}{t_a}\right)^q}\,.
\end{equation}
In both cases, the convolution operation implies that the past history of injection operations contributes to the seismicity rate at time $t$. 

The convolution operation was also used to model the seismicity rate at the Otaniemi geothermal site near Helsinki, Finland, \citep{KimA2023a}. In the model the seismicity rate was approximated as a convolution of the fluid injection rate with a kernel given by (\ref{powerlawkernel}) with  $q=2$. However, no interevent triggering was considered and the parameters of the kernel were calibrated from the decay of aftershock sequences. Specifically, the parameter $t_a$ was estimated by fitting the Omori-Utsu law to decaying event sequences between injections.

When using the Dirac $\delta$-function for the kernel $G(t)=\delta(t)$, the inducing term $\mu_\mathrm{ind}(t,\mathcal{F}_t)$ becomes proportional to the rate of fluid injection: $\mu_\mathrm{ind}(t,\mathcal{F}_t) = \mu_0 F(t - t_a)$. $t_a$ specifies a time lag from the time of the injection to the occurrence of earthquakes that takes into account the process of fluid diffusion or changes in poroelastic stresses. The unit spike response kernel assumes no memory due to past injection history. In some instances of induced seismicity, the functional form of the response function $G(t)$ may not be known in advance but can be recovered using inversion. This approach is similar to the empirical Green's function method employed in earthquake seismology.

\section{Model fitting and event forecasting methods}

\subsection{Likelihood function}

For a time dependent marked point process characterized by a set of event times and magnitudes: $\mathbf{S}=\{(t_i,m_i):\,i=1,\ldots,N\}$, the likelihood function is defined as \citep{DaleyV03a}:
\begin{equation}\label{likelihoodfun}
    L\left(\mathbf{S}|\theta,\omega\right) = e^{-\Lambda_\omega(T)}\,\prod\limits_{j=1}^{N_T}\lambda_\omega(t_j|\mathcal{H}_{t_j},\mathcal{F}_{t_j})\,
    \prod\limits_{j=1}^{N_T} f_\theta(m_j)\,,
\end{equation}

\noindent where $\lambda_\omega(t|\mathcal{H}_{t},\mathcal{F}_{t})$ is the conditional point process rate given in Eq.~(\ref{etas_induced}) and $\Lambda_\omega(T)=\int_{T_s}^{T_e} \lambda_\omega(t|\mathcal{H}_t,\mathcal{F}_t)\,dt$ is the productivity of the point process during the target time interval $[T_s,\,T_e]$ with $N_T$ number of events above a specified magnitude. $f_\theta(m)$ is the probability density function for the distribution of magnitudes. The earthquakes and fluid injection data are considered in the time interval $[T_0,\,T_e]$. This interval is subdivided into two parts: the preparatory time interval $[T_0,\,T_s[$ and the target time interval $[T_s,\,T_e]$. The events and injection data in the preparatory time interval are used to properly calibrate the conditional rate $\lambda_\omega(t|\mathcal{H}_{t},\mathcal{F}_{t})$ in the target time interval $[T_s,\,T_e]$.

It is assumed that the earthquake magnitudes follow the exponential distribution:
\begin{eqnarray}
  \label{exppdf}
  f_\theta(m) & = & \beta \exp\left[-\beta\, (m - m_0)\right]\,, \\
  F_\theta(m) & = & 1 - \exp\left[-\beta\, (m - m_0)\right]\,, \qquad \mathrm{for} \quad m\ge m_0 \,,
  \label{expcdf}
\end{eqnarray}
where parameter $\theta=\{\beta\}$ is related to the $b$-value of the Gutenberg-Richter scaling relation, $\beta = \ln(10)b$ \citep{GutenbergR54a}. $m_0$ is a prescribed lower magnitude cutoff which is specified above the catalog completeness level.

\subsection{The posterior distribution for the model parameters}

Within the Bayesian framework, the estimation of the model parameters and their uncertainties can be performed by computing the posterior distribution function. Given the magnitudes and times $\mathbf{S}_{N_T}$ of the occurrence of $N_T$ earthquakes during the target time interval $[T_s,\,T_e]$, the posterior distribution function, $p(\theta,\omega|\mathbf{S}_{N_T})$, is:
\begin{equation}\label{posterior}
    p(\theta,\omega|\mathbf{S}_{N_T}) \propto L\left(\mathbf{S}_{N_T}|\theta,\omega\right) \pi(\theta,\omega)\,,
\end{equation}
where $\pi(\theta,\omega)$ is the prior knowledge for the model parameters. The posterior distribution function updates the prior knowledge on model parameters by using the observational data through the likelihood function (\ref{likelihoodfun}).

\subsection{Bayesian predictive distribution}

For earthquake forecasting, the probability that the magnitude of the largest expected event $m_\mathrm{ex}$ will exceed a prescribed value $m$ during a future forecasting time interval $[T_e,\, T_e + \Delta T]$ is of critical importance. Within the Bayesian framework, this probability can be computed from the Bayesian predictive distribution \citep{ShcherbakovZO18a,ShcherbakovZZO19a,Shcherbakov21a}:
\begin{equation}\label{bpd}
  P_\mathrm{B}(m_\mathrm{ex} > m|\mathbf{S},\Delta T) = \int\limits_\Omega\int\limits_\Theta
  P_\mathrm{EV}(m_\mathrm{ex} > m|\theta,\omega,\Delta T)\,p(\theta,\omega|\mathbf{S})\,d\theta\,d\omega\,,
\end{equation}
where $\Theta$ and $\Omega$ are the multidimensional domains of the model parameters. $P_\mathrm{EV}(m_\mathrm{ex} > m|\theta,\omega,\Delta T)$ is the extreme value distribution for the marked point process and $p(\theta,\omega|\mathbf{S})$ is the posterior distribution (\ref{posterior}) for the model parameters.

\subsection{Seismicity forecasting}

To simulate the models during the forecasting time interval $[T_e,\,T_e+\Delta T]$, the thinning algorithm was implemented. Because the conditional rate (\ref{etas_induced}) of the ETAS models depends on the past history, all events prior to the forecasting time interval were used to calibrate the seismicity rate during the simulations.

The stochastic simulation of the ETAS model requires the specification of the model parameters. However, the true model parameters are unknown and the point estimates of the model parameters typically contain uncertainties. Therefore, for forecasting purposes, it is critical to incorporate the model parameter uncertainties into the forecasts. This can be achieved by using the Bayesian framework \citep{ShcherbakovZZO19a,Shcherbakov21a}. In the Bayesian framework, one can sample the posterior distribution of the model parameters using the Markov Chain Monte Carlo (MCMC) method to produce the chain of the model parameters (see Supplemental Material). It is also possible to examine the marginal distributions of each model parameter and assess their uncertainties. In addition, one can use the MCMC chains of the model parameters to create an ensemble of model forecasts.

For each simulation during the forecasting time interval, one can compute the number of events generated and also extract the event with the maximum magnitude. This can be done for each set of parameters from the MCMC chain generated when sampling the posterior distribution. The distribution of the maxima of events will approximate the Bayesian predictive distribution (\ref{bpd}), that can be used to compute the probabilities of having the largest expected events during the forecasting time interval $[T_e,\,T_e+\Delta T]$. The distribution of the number of events can be used to forecast the number of events above a certain magnitude \citep{ShcherbakovZZO19a,Shcherbakov21a}.

\section{The Geysers seismicity catalog and fluid injection data}\label{data}

For the application of the above formulated rate models, I considered the seismicity and water injection data from the Geysers geothermal field in California, U.S.A. (Fig.~\ref{fig1}). This includes the earthquake catalog that spans from 2003/04/30 to 2015/01/31 and the corresponding fluid injection data. During the study period the Lawrence Berkeley National Laboratory was operating a high resolution seismic network that allowed to detect and process significant number of seismic events and release the corresponding catalog \citep{MajerP2007a}. The earthquake catalog was downloaded from the Northern California Earthquake Data Center (\url{https://ncedc.org/egs/catalog-search.html}). The monthly water injection data was obtained from the California Department of Conservation (\url{https://www.conservation.ca.gov/calgem/geothermal/manual}).

\section{Results}

\subsection{Application of the stochastic rate models to the Geysers seismicity}

The above formulated model, Eq. (\ref{etas_induced}), was applied to approximate the seismicity rate at the Geysers geothermal site. Earthquakes between 2003/04/30 and 2015/01/31 and within a rectangular region shown in Fig.~\ref{fig1}a were considered. The monthly fluid injection data during the same time period (Fig.~\ref{fig1}b) was used in the model where the background term $\mu_\mathrm{ind}(t,\mathcal{F}_t)$ depended on the fluid injection rate. The stressing rate $F(t)$ was computed by normalizing the monthly fluid injection rate by the maximum rate.

\begin{figure}[!ht]
\centering
{\includegraphics*[scale=0.75, viewport= 15mm 45mm 210mm 240mm]{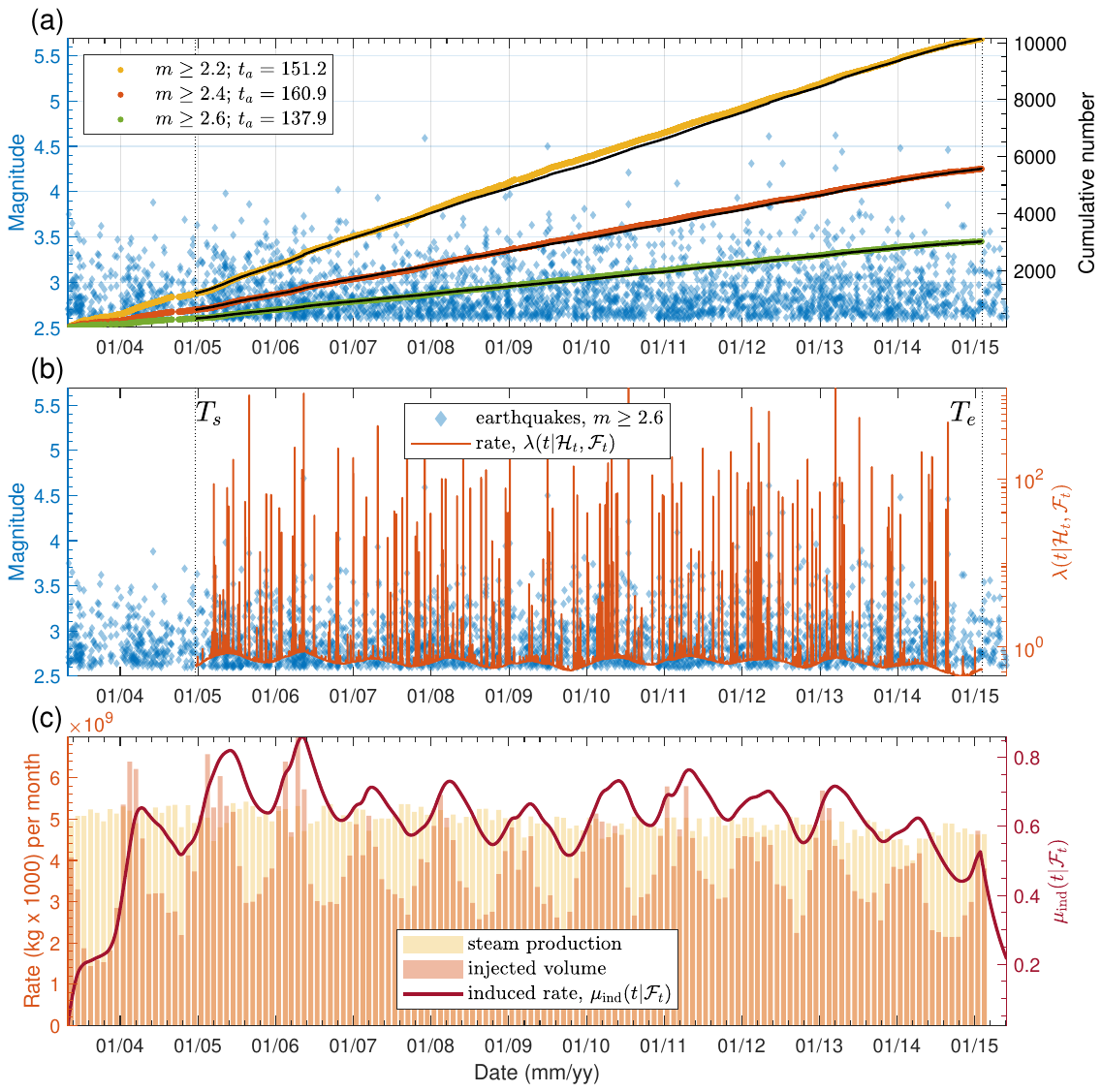}}
\caption{Fit of the stochastic process given by Eq.~(\ref{etas_induced}) to model the induced seismicity rate at the Geysers geothermal field. (a) Earthquake magnitudes above magnitude $m\ge 2.6$ are shown as blue solid diamonds during the study time interval from May 2003 to February 2015. The cumulative numbers of events above three lower magnitude cutoffs are plotted as colored dots. The fits of the stochastic model, Eq.~(\ref{etas_induced}), with the exponential response kernel, Eq.~(\ref{expkernel}), are plotted as black solid lines for each sequence with the corresponding lower magnitude cutoffs. The estimated characteristic time scales $t_a$ are given in the legend. (b) The plot of the conditional earthquake rate (\ref{etas_induced}) as fitted to the sequence with a lower cutoff of $m\ge 2.6$. (c) The estimated inducing term $\mu_\mathrm{ind}(t,\mathcal{F}_t)$ is given as a solid curve. The injection fluid and steam production volumes are shown as vertical bars.}
\label{fig2}
\end{figure}

The fits of the ETAS model using the inducing term given as a convolution operation, Eq.~(\ref{rate_conv}), with the exponential kernel, Eq.~(\ref{expkernel}), are given in Fig.~\ref{fig2}a for several lower magnitude cutoffs $m\ge 2.2$, $2.4$, and $2.6$. The fits of the other three models are reported in Figs.~S2-S4. The estimation of the model parameters was performed by using the maximum likelihood method. The summary of the estimated parameters of the four models and their performance assessed by the Akaike Information Criterion (AIC) is given in Table~\ref{table1} for earthquakes above magnitude $m\ge 2.6$. The results for other two lower magnitude thresholds ($m\ge 2.2$ and $2.4$) are given in Tables~S1-S2. Among the four considered models the ETAS models with the convolution term, Eq.~(\ref{rate_conv}), produced the best overall fit. The use of the exponential kernel, Eq.~(\ref{expkernel}), or the power-law kernel, Eq.~(\ref{powerlawkernel}), produced comparable results. However, the exponential kernel is characterized by a single characteristic time-scale parameter $t_a$ and produced more consistent fits and forecasting abilities. The parameter $t_a$ was estimated for the earthquake sequences with different lower magnitude cutoffs (Fig.~\ref{fig2}a) that characterizes a temporal response of the subsurface to external forcing with weak dependence on the lower magnitude cutoff. Fig.~\ref{fig2}b plots the full seismicity rate $\lambda_\omega(t|\mathcal{H}_t,\mathcal{F}_t)$ and the observed events above magnitude $m\ge 2.6$. The estimated induced seismicity rate $\mu_\mathrm{ind}(t,\mathcal{F}_t)$ along with the fluid injection rate and steam production are given in Fig.~\ref{fig2}c.

\begin{table*}
\tbl{The estimated parameters of the four stochastic models applied to induced seismicity at the Geysers. All earthquakes between 2003/04/30 and 2015/01/31 and above magnitude $m\ge 2.6$ were used. \label{table1}}
{\begin{tabular}{lcccccccccc}
\hline
\rule{0pt}{15pt}
\textbf{Model}         & $\boldsymbol K$   & $\boldsymbol c$    & $\boldsymbol p$  & $\boldsymbol \alpha$ & $\boldsymbol \mu_0$ & $\boldsymbol t_a$ & $\boldsymbol q$ & \textbf{AIC} \\[1mm]
\hline
\rule{0pt}{15pt}
ETAS          & 60.2 & 0.00016 & 1.20 & 2.27 & 0.64 &  --   &  --  & 5784.71\\[1mm]
\hline
\rule{0pt}{15pt}
ETAS+frac     & 69.4 & 0.00011 & 1.10 & 2.22 & 1.09 & 2.8   &  --  & 5810.75\\[1mm]
\hline
\rule{0pt}{15pt}
ETAS+conv(exp)& 58.9 & 0.00016 & 1.22 & 2.28 & 4.92 & 137.9 &  --  & 5754.28\\[1mm]
\hline
\rule{0pt}{15pt}
ETAS+conv(pl) & 58.7 & 0.00017 & 1.22 & 2.28 & 0.05 & 145.2 & 2.53 & 5753.72\\[1mm]
\hline
\end{tabular}}
{}
\end{table*}

\subsection{Forecasting the earthquake numbers and the largest expected events}

It is important to assess the forecasting abilities of the proposed stochastic models. For the retrospective forecasting, I considered progressively increasing training time intervals and used a fixed forecasting time interval of $\Delta T = 120$ days. Specifically, the starting date was set at 2003/04/30, which corresponded to $T_0=0$ with $T_s=600$ days. The end of the training time interval, $T_e$, was progressively shifted forward by half a year starting from 2010/01/01. For the forecasting purpose, I used the ETAS model with the inducing term (\ref{rate_conv}) and exponential kernel (\ref{expkernel}). To forecast the evolution of the sequence during each forecasting time interval $[T_e,\,T_e+\Delta T]$, the model was simulated 50,000 times using the parameters from the MCMC chain generated when sampling the posterior distribution in the target time interval $[T_s,\,T_e]$. This allowed me to incorporate fully the model parameter uncertainties into the forecasts (see Supplemental Material Figs.~S5-S7).

\begin{figure}[!ht]
\centering
{\includegraphics*[scale=0.75, viewport = 15mm 45mm 210mm 240mm]{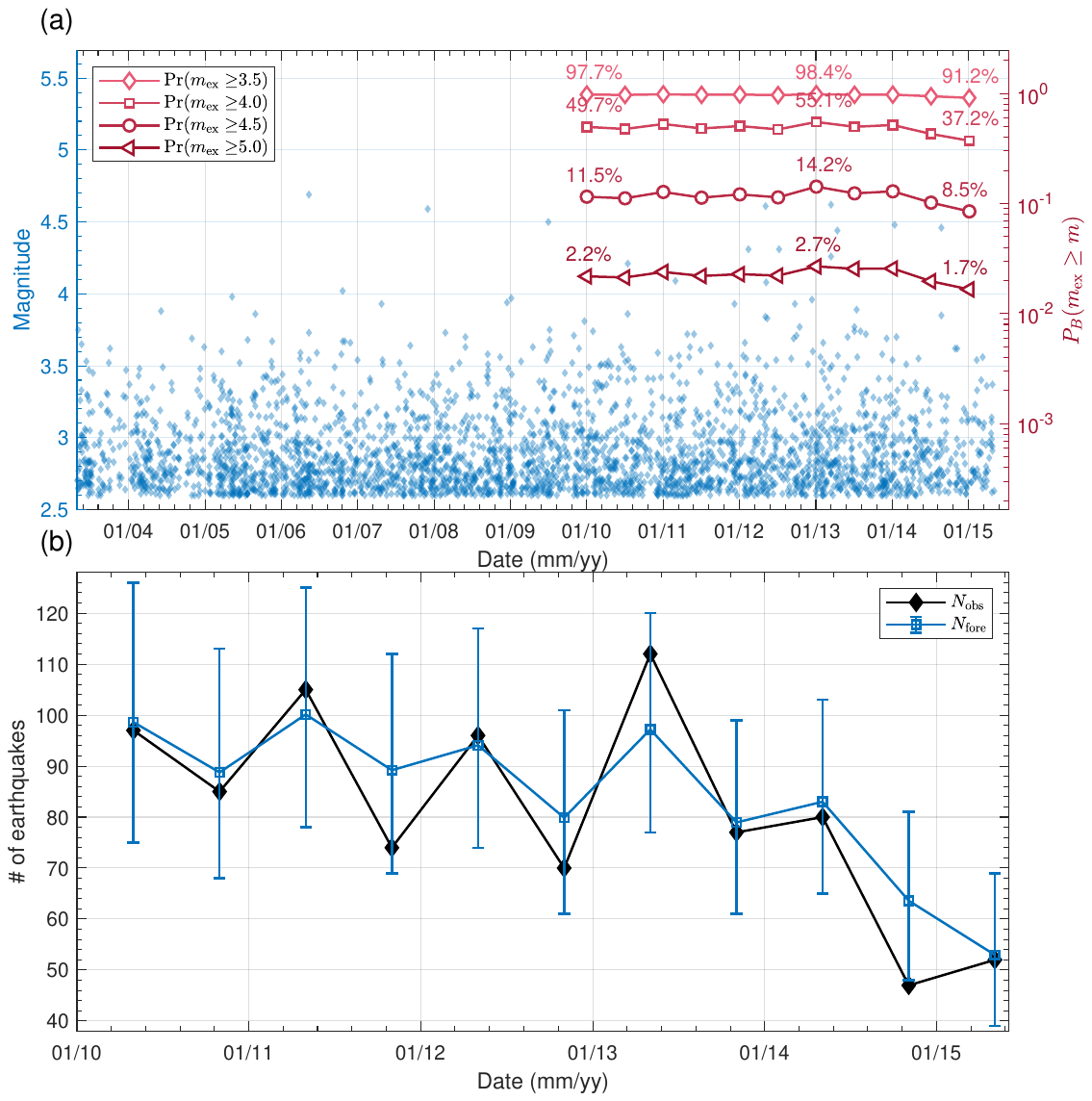}}
\caption{Earthquake forecast probabilities and the comparison of the observed and forecasted numbers of events. (a) The probabilities for the occurrence of the largest expected events above magnitudes $m_\mathrm{ex}\ge 3.5$, $4.0$, $4.5$, and $5.0$ are plotted as open symbols. The forecasting time interval is $\Delta T = 120$ days. The forecasts are given starting from January 2010 and updated every 183 days. (b) The observed and forecasted numbers of events above magnitude $m\ge 2.6$ in each forecasting time interval starting from January 2010.}
\label{fig3}
\end{figure}

For the retrospective analysis of the Geysers seismicity, the average number of forecasted events with the corresponding 95\% confidence bounds are plotted in Fig.~\ref{fig3}b along with the observed number of events in each forecasting time interval for magnitudes above $m\ge 2.6$. The model accurately forecasted within uncertainty bounds the observed events for the most of the forecasting time intervals. It slightly overestimated the forecasted number of events for the forecasted time interval ending on 2014/11/03. It also followed the trend in the observed numbers in contrast to the standard ETAS model with the constant background rate $\mu_0$ (Fig.~S8).

In addition, the Bayesian predictive distribution was computed from the distribution of maximum magnitudes extracted from each simulation run for the specified forecasting time intervals \citep{Shcherbakov21a}. This was used to compute the probabilities for having the largest expected events during each forecasting time interval and above magnitudes $m_\mathrm{ex}\ge 3.5$, $4.0$, $4.5$, and $5.0$ (Fig.~\ref{fig3}a). As expected the probabilities decrease for larger expected events. The probabilities also depend on the duration of the forecasting time interval $\Delta T$ and increase with its length. 

The implemented approach can be employed for prospective forecasting. The past seismicity and history of injection operations can be used to sample the posterior distribution of the model parameters to generate the corresponding MCMC parameter chains during the past training time interval. Next, one needs to specify suitable scenarios for injection operations during the future forecasting time interval $[T_e,\,T_e+\Delta T]$. By simulating the stochastic model, Eq.~(\ref{etas_induced}), forward in time in the interval $[T_e,\,T_e+\Delta T]$, one can generate an ensemble of earthquake catalogs from which the Bayesian predictive distribution can be constructed and the corresponding probabilities for the largest expected events computed. This can be used to quantify the effect of fluid injection scenarios on the occurrence of the largest events.

\subsection{Forecast testing}

\begin{figure}[!ht]
\centering
{\includegraphics*[scale=0.75, viewport= 15mm 65mm 200mm 208mm]{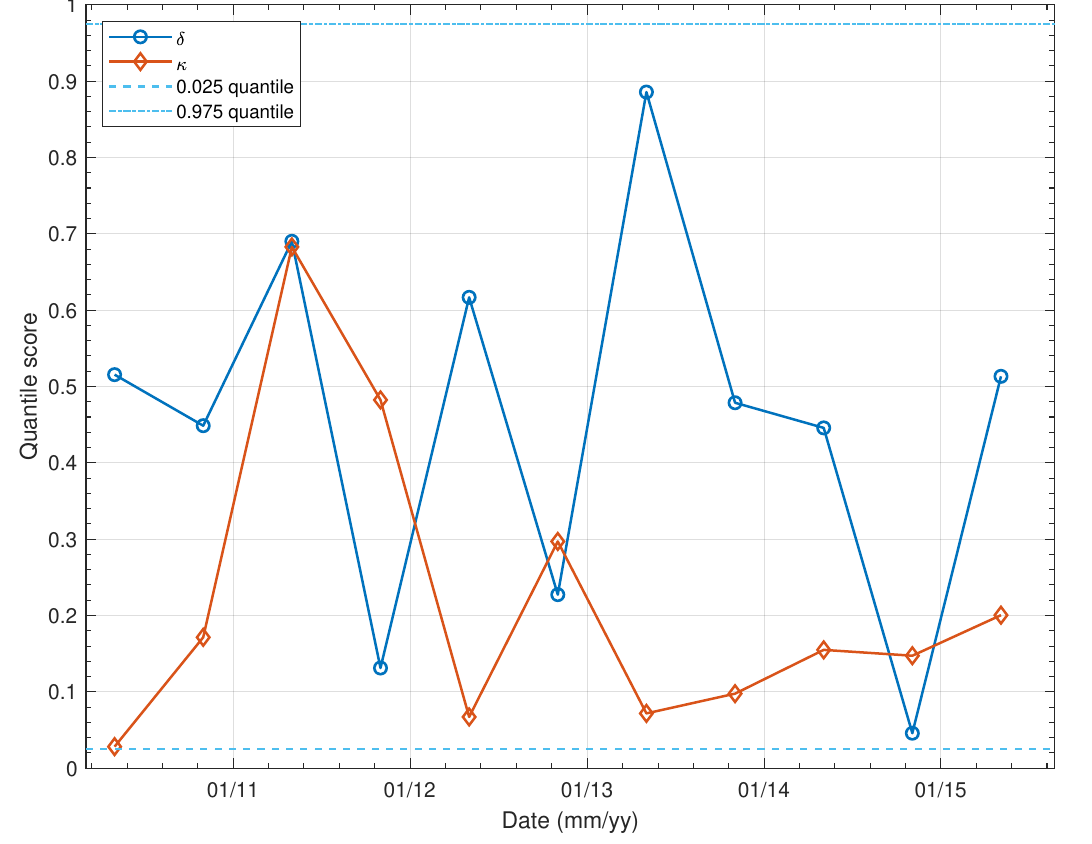}}
\caption{Plots of the quantile scores $\delta$ (for N-test) and $\kappa$ (for M-test). The stochastic model, Eq.~(\ref{etas_induced}), with the exponential response function, Eq.~(\ref{expkernel}), assuming a single characteristic time scale $t_a$ in the inducing term Eq.~(\ref{rate_conv}) was used. The scores were computed at the end of each forecasting time interval. }
\label{fig4}
\end{figure}

The performance of the model during the forecasting time intervals can be evaluated by applying several statistical tests (see Supplemental Material). I applied the N-test to assess how well the model forecasted the number of events above magnitude $m\ge 2.6$ during each forecasting time interval. The plot of the quantile score $\delta$ is given in Fig.~\ref{fig4}. It is assumed that the values between the lower quantile of $0.025$ and upper quantile of $0.975$ indicate passing the test. Otherwise, the forecast either overestimates or underestimates the observed number of events. The results show that the forecasts were accurate for all of the forecasting time intervals. To assess the consistency of the magnitude distribution of the forecasted events, the M-test was applied. This test is characterized by the quantile score $\kappa$. Similarly to the N-test, the quantile score $\kappa$ was found between lower and upper quantile values (Fig.~\ref{fig4}) indicating a consistent reproduction of the frequency-magnitude distribution of the observed magnitudes. For comparison, the earthquake forecasts and forecast testing using the original ETAS model with constant background rate are given in Figs.~S8 and S9. The results indicate that the model with the constant background rate does not forecast well the observed number of events during the forecasting time intervals.

\section{Discussion and conclusions}\label{discus}


The original ETAS model was formulated primarily to account for the occurrence of aftershocks and typically assumes a constant background rate. However, seismic activity, associated with earthquake inducing anthropogenic processes, signifies that other physical mechanisms can be at play when triggering earthquakes \citep{GrigoliCPR17a,SchultzSBEBE20a}. In many cases, these triggering aspects are associated with the subsurface fluid injection. As a result, the corresponding redistribution of poroelastic and/or thermal stresses and changes in fluid pore pressure can induce earthquakes. Therefore, it is important to incorporate these anthropogenic factors into stochastic models that are used in modelling induced seismicity.

The brittle layer of the Earth, where earthquakes nucleate, propagate and interact, can be treated as a nonequilibrium thermodynamic system. The response of this system to weak external forcing can be considered, to a first order, as linear. In this work, I hypothesize that triggering of induced earthquakes is the result of such external forcing. To quantify this triggering, a mechanism has been added to the conditional earthquake rate, Eq.~(\ref{etas_induced}), to quantify the delay in the occurrence of earthquakes and to condition on the past history of forcing. The background term, that uses the convolution operation, Eq.~(\ref{rate_conv}), assumes that the effect of the changes in the fluid injection rate on seismicity is the result of the linear response of the viscoelastic medium combined with the past history of injection. The kernel, Eq.~(\ref{expkernel}) or Eq.~(\ref{powerlawkernel}), plays the role of an impulse-response function of the medium to a $\delta$-like injection rate spike. 

Several physical mechanisms have been suggested to explain the occurrence of induced earthquakes \citep{ShapiroD09a,SegallL15a,BhattacharyaV19a}. However, their incorporation into stochastic rate models for probabilistic earthquake forecasting is not always straightforward. It is reasonable to assume that a number of mechanisms, operating at the same time and to a certain degree, contribute to the occurrence of induced events and may operate at different time scales. In this respect, linear response theory offers a rather direct way of quantifying the effects of fluid injection operations and formulating a stochastic rate model that takes into account the effective rheological properties of the subsurface. In physics, linear response theory has been instrumental in estimating the material parameters such as the magnetic susceptibility or the dielectric functions of physical systems \citep{KuboTH1991a,LiviP2017a}. It is actively used in climate studies, for example, to relate the global mean temperature to changes in the CO$_2$ concentration in the atmosphere \citep{LucariniBHRPW2014a,LemboLR2020a}. It is also used in engineering, neurophysiology, signal theory, control theory, etc. \citep{Gottwald2020a}. For these systems the internal dynamics can be nonlinear but the response to weak external perturbations appears to be linear in nature.

The functional form of the kernel $G(t)$ reflects the physical response of the underlying system to external forcing \citep{HasselmannHGOS1997a,Lucarini2018a}. It is characterized by the presence of characteristic time scale(s) that are inherent to the physics of the system. Within the framework of linear response theory, the occurrence of induced earthquakes can be treated as a response of the subsurface system to external perturbations. The subsurface is characterized by strong material heterogeneity and complex stress perturbations, which are modulated by the subsurface fluid injections. As a result, the characteristic time scale reflects the effective rheological properties of the subsurface. For example, the exponential form (\ref{expkernel}) of the response function $G(t)$ assumes that Maxwell's rheology is applicable, where characteristic time scale $t_a$ is proportional to the effective viscosity of the subsurface \citep{BurgmannD08a}. This is consistent with the study of the damped harmonic oscillator for which the exponential functional form of $G(t)$ was derived analytically \citep{LiviP2017a}. In the model the characteristic time-scale parameter, that appears in the exponential term, is proportional to the damping coefficient and inversely proportional to the stiffness of the spring. For a Maxwell viscoelastic solid, the effective viscosity is related to the characteristic time scale as $\eta=\mu\,t_a$, with the average estimated value of $t_a\approx 150$ days and the shear modulus of $\mu=25$ GPa, this gives $\eta\approx 3.2\times 10^{17}$ Pa$\cdot$s. Geodetic observations and geomechanical modelling give a range of values of $10^{16}$--$10^{20}$ Pa$\cdot$s for the effective viscosity of the crust \citep{BurgmannD08a}. It critically depends on the temperature regime, crustal heterogeneities, presence of fluids and other factors. For volcanic and geothermal systems the effective viscosity values of $10^{16}$--$10^{19}$ Pa$\cdot$s are considered when modelling the surrounding rocks \citep{NewmanDOD2001a,HeadHGF2021a}. Therefore, the obtained value for the effective viscosity is plausible for the Geysers geothermal site.

On the other hand when using a power-law like response function (\ref{powerlawkernel}), it is assumed that a nonlinear power-law rheology can be at play \citep{FreedB04a}. One possible manifestation of such rheological behaviour is the occurrence of aftershocks \citep{ZhangS16a}. The decay rate of aftershocks is typically approximated by the Omori-Utsu empirical law. It was shown by \cite{ZhangS16a}, when analyzing a slider-block model with nonlinear viscoelastic coupling, that the stress transfer rate in the model is, $\frac{d\sigma}{dt}\propto \frac{d}{dt} \left[t^{-\frac{1}{n-1}}\right] \propto \left[t^{-\left(1+\frac{1}{n-1}\right)}\right]$, where $n$ is a power-law exponent of the nonlinear viscoelastic rheology. Assuming that the response function $G(t)$, Eq.~(\ref{powerlawkernel}), has a similar functional form as the stress transfer rate one gets, $q = 1 + \frac{1}{n-1}$. From the estimated value of $q = 2.53$ one can obtain the power-law exponent $n = 1 + \frac{1}{q-1} = 1.7$. This shows that nonlinear rheological effects are not that strong and Maxwell's rheology dominates the behavior. This is also reflected in the comparative fits of the both models to the observed seismicity rate. In fact, the both kernels have a similar functional shape when using the estimated parameters (Fig.~S10).

In summary, a stochastic model based on the ETAS process was introduced to approximate the seismicity rate at the Geysers geothermal site. In the suggested approach, the background term was formulated as a convolution operation between the fluid injection rate and a specific kernel (response function). The convolution operation can be justified by invoking linear response theory. This theory is extensively used in nonequilibrium statistical mechanics to describe the effects of forcing on out of equilibrium thermodynamical systems. By employing the Bayesian predictive framework, the stochastic earthquake rate model was used to compute retrospectively the probabilities for the occurrence of the largest expected events during the evolution of the sequence and above a certain magnitude. The validity of these forecasts were confirmed by applying several statistical tests. In addition to modeling the seismicity rate, this study provides a compelling evidence that external forcing, such as fluid injection, can be used for probing the transient rheological properties of the subsurface on longer time-scales in contrast to the ones associated with seismic wave propagation.

\begin{datres}
The earthquake catalog was downloaded from the Northern California Earthquake Data Center (\url{https://ncedc.org/egs/catalog-search.html}). The monthly water injection data was obtained from the California Department of Conservation (\url{https://www.conservation.ca.gov/calgem/geothermal/manual}).

The Matlab software codes developed to perform the analysis are freely available at Zenodo (\url{https://zenodo.org/doi/10.5281/zenodo.10936424}).
\end{datres}

Supplemental Material provides additional details for the methods used in the analysis. The estimated model parameters are provided in Tables S1-S2. Additional results are plotted in Figures S1-S10.

\section{Declaration of Competing Interests}

The author acknowledges that there are no conflicts of interest recorded.\footnote{The author acknowledges that there are no conflicts of interest recorded.}

\begin{ack}
This research has been supported by the NSERC Discovery grant. Constructive criticism and useful comments by David Dempsey, one anonymous reviewer, and Associate Editor, Maximilian Werner, helped to clarify the results and improve the presentation.
\end{ack}

\vspace{20mm}
\noindent Robert Shcherbakov$^{1,2\ddag}$ \\
\\
$^{1}$Department of Earth Sciences, Western University, London, Ontario, \emph{N6A 5B7}, Canada\\
$^{2}$Department of Physics and Astronomy, Western University, London, Ontario, \emph{N6A 3K7}, Canada\\
\\[5mm]
$^\ddag$E-mail: rshcherb@uwo.ca

\end{document}